\documentstyle[preprint,aps,pre]{revtex}
\def\be{\begin{equation}}
\def\ee{\end{equation}}
\def\bea{\begin{eqnarray}}
\def\eea{\end{eqnarray}}
\begin{document}
\title{Linearizability of the Perturbed Burgers Equation}
\author{R. A. Kraenkel, J. G. Pereira and E. C. de Rey Neto}
\vskip 0.5cm
\address{Instituto de F\'{\i}sica Te\'orica\\
Universidade Estadual Paulista\\
Rua Pamplona 145\\
01405-900\, S\~ao Paulo \\ 
Brazil}
\maketitle
\begin{abstract}
We show in this letter that the perturbed Burgers equation 
$$ 
u_t = 2uu_x + u_{xx} + \epsilon \left( 3 \alpha_1 u^2 u_x + 
3\alpha_2 uu_{xx} + 3\alpha_3 u_x^2 + \alpha_4 u_{xxx} \right)  
$$ 
is equivalent, through a near-identity transformation and up to 
$\cal{O} (\epsilon)$, to a linearizable equation if the condition 
$3\alpha_1 - 3\alpha_3 - \frac{3}{2}\alpha_2 + \frac{3}{2}\alpha_4
= 0$  is satisfied. In the case this condition is not
fulfilled, a normal form for the equation under consideration is
given. Then, to illustrate our results, we make a linearizability
analysis of the equations governing the dynamics of a
one--dimensional gas.  
\end{abstract}

\vskip 1.0cm


\vskip 1.0cm

\newpage

\section{Introduction}

The object of this letter is the perturbed Burgers equation
\begin{equation} 
u_t = 2uu_x + u_{xx} + \epsilon \left( 3 \alpha_1 u^2 u_x +
3\alpha_2 uu_{xx} + 3\alpha_3 u_x^2 + \alpha_4 u_{xxx} \right)
\quad , 
\label{principal} 
\end{equation} 
where $\alpha_i$ are constants, $\epsilon \ll 1$ is a 
perturbative parameter, and  subscripts denote partial
differentiation. It appears in the long-wave, small amplitude
limit of extended systems dominated by dissipation, but where
dispersion is also present at a higher order. More precisely, 
those systems described by equations whose linear part admits a
dispersion relation of  the form
\begin{equation}
\Omega \left( k \right) =  a_3 k^3 + a_5 k^5 + \cdots + \imath
\left( b_2 k^2 + b_4 k^4+ 
\cdots \right) \quad ,
\label{disper}
\end{equation}
with $a_i$ and $b_i$ real constants.
For example, Eq.(\ref{principal}) appears in the description of
gas  dynamics~\cite{dodd}, and in certain cases of 
free-surface motion of waves in heated fluids~\cite{kpm1}. More
important, however, is  the fact that the terms
appearing at order $\epsilon$ are the only ones allowed if
Eq.(\ref{principal}) is obtained from long-wave perturbation
theory, and no constants are allowed to scale with $\epsilon $.
In this sense, it has the {\it universality} characteristics,
much in the same way as the equations discussed by Calogero
\cite{calo}.

When the $\cal{O} (\epsilon )$ terms are discarded, we have
simply Burgers equation, which is an equation linearizable
through a Hopf--Cole transformation~\cite{whit}. It is, thus, a
natural question to ask whether Eq.(\ref{principal}) is also
linearizable. Put in  this way, the answer is that it is
linearizable if $ \alpha_1 =  \alpha_2 = \alpha_3 = 
\alpha_4 $, in which case the equation is reduced to the sum of
Burgers with  the first higher--order equation of the Burgers
hierarchy~\cite{hier}. We notice in passing that the latter is
also linearizable by the same Hopf--Cole transformation that
linearizes the Burgers equation.  However, we  can put the
question on a more general setting by introducing the idea of
{\it near identity transformation} \cite{arno}, that is, a
transformation $ u \rightarrow w$ of  the form 
\begin{equation}
 u =  w + \epsilon \phi \left( w \right) \quad .
 \label{near}
\end{equation}
If we apply such a transformation to Eq.(\ref{principal}), we
may look for  functions $ \phi \left( w \right) $ such that the
transformed equation reads:
\begin{equation}
w_t = 2ww_x + w_{xx} + \epsilon \lambda \left( 3w^2 w_x +
3ww_{xx} + 3w_x^2 + w_{xxx} \right) + {\cal{O}}\left( {\epsilon}
^2 \right) \quad , 
\label{tprincipal}  
\end{equation}
for some $\lambda \in R $. If such a $\phi \left( w \right)$
exits, we say that  Eq.(\ref{principal}) and 
Eq.(\ref{tprincipal}) are equivalent up to order $\epsilon $. As
Eq.(\ref{tprincipal}) is linearizable, so is Eq.(\ref{principal})
up to $\cal{O} (\epsilon )$. The 
fundamental issue here is thus to determine the conditions for
the existence of a near identity transformation (that is, $\phi
\left( w \right)$ ) ensuring the equivalence, up to $\cal{O}
(\epsilon )$, of equations (\ref{principal}) and
(\ref{tprincipal}).  This is the question we will address in this
letter, and an answer will be given in terms of a condition on the
parameters $\alpha_1$, $\alpha_2$, $ \alpha_3 $ and $\alpha_4 $.

The kind of equivalence defined above has been introduced in
ref.~\cite{kodama1} in the context of asymptotic evolution
equations. For dispersive systems whose lowest order, in the 
long-wave approximation, is described by the Korteweg-de Vries
(KdV) equation, it has  been shown that there exists always a
near-identity transformation  which makes the $\cal{O}(\epsilon
)$ perturbation integrable. The same is true for the case of the 
Nonlinear Schr\"odinger equation \cite{kodama2}. However,
obstacles to integrability  appear in ${\cal{O}}({\epsilon}^2 )$
\cite{kodama2,bbm}. The same kind of ideas has been  used in 
\cite{fokas} to show that, in the KdV case, a $ \phi \left( w
\right)$ depending  explicitly on $x$ can be found as to
completely remove the $\cal{O}(\epsilon )$  correction. We will
come back to this issue at the end of this letter. 

We will show that, in the case of Eq.(\ref{principal}), obstacles
to linearizability  appear already at $\cal{O}(\epsilon )$. We
mean by this that, in general, Eq.(\ref{principal}) is not
equivalent to Eq.(\ref{tprincipal}). The condition for the
equivalence will be shown to be 
\begin{equation}
3\alpha_1 - 3\alpha_3 - \frac{3}{2}\alpha_2 +
\frac{3}{2}\alpha_4 = 0 \quad .
\label{condicao}
\end{equation}
Furthermore, in the case where condition (\ref{condicao}) is not
satisfied, we find a  normal form for Eq.(\ref{principal}), that
is, a form to which Eq.(\ref{principal})  can always be
transformed. Finally, as an illustration, we make a
linearizability analysis of the equations governing the dynamics
of a one--dimensional gas, and we show that, already at order
${\cal O}(\epsilon)$, these equations can not be linearized. 

\section{Linearizability Analysis}

Let us then implement the ideas exposed above. We want to insert 
Eq.(\ref{near}) into Eq.(\ref{principal}), discard all
${\cal{O}}({\epsilon}^2 )$ terms, and compare the result with
Eq.(\ref{tprincipal}). To do so, we have to specify the possible
form of $\phi \left( w \right)$. They ought to be such as to
generate, at $\cal{O}(\epsilon )$, terms  of the form $ w^2 w_x$,
$ww_{xx}$, $w_x^2$ and $w_{xxx}$. The allowable terms turn out  to
be $ w_x$, $w^2$ and $w_x \partial ^{-1}w$, where $\partial ^{-1}$
means integration  in $x$. Thus the general form of $\phi \left( w
\right)$ is :
\begin{equation}
\phi \left( w \right) = \alpha w_x + \beta w^2 + \gamma w_x
\partial ^{-1}w \quad ,
\label{phi}
\end{equation}
 where $\alpha$, $\beta$ and $\gamma$ are constants to be
determined. 
 
We introduce now the following useful notations:
\begin{equation}
B \left( u \right) = 2uu_x + u_{xx} \quad ,
\label{b}
\end{equation}
and
\begin{equation}
\Theta\left( u \right) = 3 \alpha_1 u^2 u_x + 3\alpha_2 uu_{xx} +
3\alpha_3 u_x^2 + \alpha_4 u_{xxx} \quad .
\label{teta}
\end{equation}
Accordingly,  Eq.(\ref{principal})  becomes 
\begin{equation}
u_t =  B\left( u \right) + \epsilon \Theta\left( u \right) \quad .
\label{novaeq}
\end{equation}
The transformation (\ref{near}) changes Eq.(\ref{novaeq}) to an
equation in $w$, given by
\begin{equation}
w_t = B\left( w \right) + \epsilon \left\{ \Theta\left( w \right)
+ \left[ B\left( w \right),\phi \left( w \right) \right] \right\}
+ {\cal{O}}\left( {\epsilon} ^2 \right) \quad ,
\label{weq}
\end{equation}
where 
$$ \left[ B\left( w \right),\phi \left( w \right) \right] =
\frac{\delta B}{\delta w} \phi - \frac{\delta \phi}{\delta w} B
\quad .
$$
In order to obtain the transformed equation, we have thus to
calculate the commutator $\left[ B\left( w \right),\phi \left( w
\right) \right] $, which is the tedious part of  our task. After
performing that calculation, we get
\begin{equation}
\left[ B\left( w \right),\phi \left( w \right) \right] =   2\beta   w_x^2 + 
 2\gamma   ww_{xx} + \left( 2\beta + \gamma \right) w^2 w_x \quad.
\label{comuta}
\end{equation}
Inserting this into Eq.(\ref{weq}), the transformed equation
reads:
\begin{eqnarray}
w_t = B \left( w \right) + \epsilon \left[ \left( 2\beta + \gamma
+3\alpha_1 \right)  w^2w_x \right. &+& \left( 2\gamma + 3\alpha_2
\right) ww_{xx} \nonumber \\  &+& \left. \left( 2\beta + 3
\alpha_3 \right) w_x^2 + \alpha_4 w_{xxx} \right] \quad .
\label{ttequation}
\end{eqnarray}
If we now require Eq.(\ref{ttequation}) to be of the form given
by Eq.(\ref{tprincipal}), we have to take $\lambda = \alpha_4 $,
and the following conditions must be satisfied:
\begin{equation}
2\beta = 3\alpha_4 - 3\alpha_3 \quad ,
\label{cond1}
\end{equation}
\begin{equation}
2\gamma = 3\alpha_4 - 3\alpha_2 \quad ,
\label{cond2}
\end{equation}
\begin{equation}
2\beta + \gamma = 3\alpha_4 - 3\alpha_1 \quad .
\label{cond3}
\end{equation}
Clearly, this system of equations is not always solvable. The
solubility condition is 
\begin{equation}
3\alpha_1 - 3\alpha_3 -\frac{3}{2}\alpha_2 + \frac{3}{2}\alpha_4
= 0 \quad ,
\label{sol}
\end{equation}
in which case $ \beta = \frac{3}{2}(\alpha_4 -\alpha_3 )$ and $
\gamma = \frac{3}{2}  (\alpha_4 - \alpha_2 ) $. Note that $\alpha
$ is left undetermined. Condition  (\ref{sol}) is thus the
condition that must be satisfied in order to make 
Eq.(\ref{principal}) equivalent, up to $\cal{O} (\epsilon )$, to 
Eq.(\ref{tprincipal}).

Suppose now that Eq.(\ref{sol}) is not satisfied. The general
form of the transformed  equation is given by
Eq.(\ref{ttequation}). The ${\cal{O}}({\epsilon} )$ terms can be 
written as the sum of a linearizable term proportional to
$\alpha_4 F_3\left( w \right)$, with
\begin{equation}
F_3\left( w \right) = 3w^2 w_x + 3ww_{xx} + 3w_x^2 +  w_{xxx}
\quad ,
\label{f3}
\end{equation} 
plus a term $Z\left( w \right)$ representing the obstacle to
linearizability, that is,
\begin{equation}
w_t = B\left( w \right) + \epsilon \alpha_4 F_3\left( w \right)
+\epsilon Z\left( w \right) \quad ,
\label{normal}
\end{equation}
where
\begin{eqnarray}
Z\left( w \right) = \left( 2\beta + 3\alpha_3 - 3 \alpha_4
\right) w_x^2 &+& \left(  2\gamma +3\alpha_2 - 3\alpha_4 \right)
ww_{xx} \nonumber \\  &+& \left( 2\beta +\gamma + 3\alpha_1 - 
3\alpha_4 \right) w^2w_x \quad .
\label{zw}
\end{eqnarray}
If we call each of the coefficients appearing in the obstacle
respectively by $\mu_1$, $\mu_2$ and 
$\mu_3$, and if we further introduce $\nu_i$ through $\mu_i = \mu
\nu_i$,  with $ \mu = 3\alpha_1 - 3\alpha_3 -\frac{3}{2}\alpha_2 +
\frac{3}{2}\alpha_4$,  then  we may write out the {\it normal form
}of Eq.(\ref{principal}) as: 
\begin{equation}
w_t = B\left( w \right) + \epsilon \alpha_4 F_3\left( w \right)
+\epsilon \mu \left( 
\nu_1 w_x^2 + \nu_2 ww_{xx} + \nu_3 w^2 w_x \right)  \quad ,
\label{formanormal}
\end{equation}
where $\nu_i$ are arbitrary constants satisfying 
\begin{equation}
\nu_1 - \nu_3 + \frac{\nu_2}{2} = -1 \quad .
\label{eqnu}
\end{equation}
Equation (\ref{formanormal}) encompasses the main results of this
letter: for $\mu = 0$ we  have a linearizable equation, and for
$\mu \neq 0 $  it gives the  general form to which
Eq.(\ref{principal}) is equivalent up to $\cal{O} (\epsilon)$.

\section{Gas Dynamics}

Let us consider the equations governing the dynamics of a
one--dimensional gas~\cite{dodd}
\be
\rho_t + \left(\rho u \right)_x = 0 \quad ,
\label{gas1}
\ee
\be
\left(\rho u \right)_t + \left[\rho u^2 + P - \mu u_x \right]_x =
0 \quad ,
\label{gas2}
\ee
where $\rho(x,t)$ is the density, $u(x,t)$ is the velocity, $\mu$
is the viscosity, and
$$
P = A \left(\frac{\rho}{\rho_0} \right)^{\gamma} \quad ,
$$
is the pressure, with $\gamma=(c_p/c_v)$ the ratio of specific
heats, and $A$ a  proportionality constant. In order
to study its long--wave, small amplitude limit, we define slow
space and time variables,
\bea
\xi &=& \epsilon (x - c t) \quad , \label{xi} \\
\tau &=& \epsilon^2 t \quad , \label{tau}
\eea
and scale the original (primed) density and velocity fields
according to
\bea
\rho^{\prime} &=& \rho_0 + \epsilon \rho \quad , \label{ro} \\
u^{\prime} &=& \epsilon u \quad . \label{u}
\eea
In terms of these new variables, Eqs.(\ref{gas1}) and
(\ref{gas2}) becomes
\be
\rho_0 u_\xi - c \rho_\xi + \epsilon \Big[ \rho_\tau + (u
\rho)_\xi 
\Big] = 0 \quad ,
\label{gas3}
\ee
\bea
\frac{A \gamma}{\rho_0} \, \rho_\xi - c \rho_0 u_\xi + \epsilon
\Big[ - c (u \rho)_\xi + \rho_0 u_\tau + \rho_0 u_\tau 
&+& \rho_0 (u^2)_\xi - \mu u_{\xi\xi} \Big] \nonumber \\  
&+& \epsilon^2 \Big[ (u \rho)_\tau + (\rho u^2)_\xi \Big] = 0
\quad .
\label{gas4}
\eea
Moreover, as a compatibility condition at order ${\cal
O}(\epsilon^0)$, we have to set
\be
c^2 = \frac{A \gamma}{\rho_0} \quad .
\label{c2}
\ee

Now, from Eq.(\ref{gas3}) we obtain
\be
\rho_\xi = \frac{\rho_0}{c} \, u_\xi + \frac{\epsilon}{c} \left[
\rho_\tau + (u \rho)_\xi \right] + {\cal O}(\epsilon^2) \quad ,
\label{roxi}
\ee
or 
\be
\rho = \frac{\rho_0}{c} \left[u + \frac{\epsilon}{c} \left(u^2 -
\partial^{-1} u_\tau \right) \right] + {\cal O}(\epsilon^2) \quad ,
\label{ro2}
\ee
with $\partial^{-1}$ indicating an integration in the $\xi$
coordinate. Substituting in Eq.(\ref{gas4}), and using the
resulting equation into itself, we are lead to
\be
u_\tau = - u u_\xi + \frac{\mu}{2 \rho_0} \, u_{\xi\xi} + \epsilon
\left[ \frac{1}{c} u^2 u_\xi - \frac{3 \mu}{2 c \rho_0} u
u_{\xi\xi} - \frac{\mu}{4 c \rho_0} (u_\xi)^2 + \frac{\mu^2}{8 c
{\rho_0}^2} u_{\xi\xi\xi} \right] + {\cal O}(\epsilon^2) \quad .
\label{utau1}
\ee

In order to compare to Eq.(\ref{principal}), we have first to
rewrite Eq.(\ref{utau1}) in a nondimensional form. To this end, we
nondimensionalize all variables according to
\be
u \longrightarrow \frac{u}{c} \quad ; \quad \xi \longrightarrow -
\frac{\rho_0}{c \mu} \, \xi \quad ; \quad \tau \longrightarrow
\frac{\rho_0}{2 c^2 \mu} \, \tau \quad .
\label{nondi}
\ee
In terms of these new variables, the nondimensional version of 
Eq.(\ref{utau1}) reads
\be
u_\tau = 2 u u_\xi + u_{\xi\xi} + \epsilon
\left[- \frac{2 \rho_0}{A \gamma} u^2 u_\xi - \frac{3 \rho_0}{A
\gamma}  u u_{\xi\xi} - \frac{\rho_0}{2 A \gamma} (u_\xi)^2 -
\frac{\rho_0}{4 A \gamma} u_{\xi\xi\xi} \right] + {\cal
O}(\epsilon^2) \quad .
\label{utau2}
\ee   
A comparison with Eq.(\ref{principal}) yields
\be
\alpha_1 = - \frac{2 \rho_0}{3 A \gamma} \;\; ; \;\;
\alpha_2 = - \frac{\rho_0}{A \gamma} \;\; ; \;\;
\alpha_3 = - \frac{\rho_0}{6 A \gamma} \;\; ; \;\;
\alpha_4 = - \frac{\rho_0}{4 A \gamma} \;\; .
\label{alphas}
\ee
The linearizability condition (\ref{sol}), therefore, is
\be
\frac{3 \rho_0}{8 A \gamma} = 0 \quad .
\ee
This means that, in the long--wave, small amplitude limit, the
equation governing a one--dimensional gas can be linearized only
at the lowest order. When the ${\cal O}(\epsilon)$ corrections are
taken into account, the corresponding equation can not be
linearized, indicating that obstacles to linearizability are
present already at this order.

\section{Final Remarks}

The first remark is about the traveling--wave solution to the 
Burgers equation. Suppose that we want to know the $\cal{O}
(\epsilon)$ correction to the solution 
\begin{equation}
w = -k\left[ 1 - \tanh \left( kx - 2k^2 t \right) \right]
\label{kink}
\end{equation}
of the Burgers equation. What is remarkable about it is that we
may take $Z\left(  w \right) = 0$. That is, there exist constants
$\nu_i$, namely $\nu_1 = -\nu_2  = -\nu_3 = -\frac{2}{3}$,
satisfying condition (\ref{eqnu}) such that the obstacle  term in
Eq.(\ref{formanormal}) is equal to zero. This makes it easy to
find the $\cal{O} (\epsilon)$ correction to the solution
(\ref{kink}) from the joint solution of Burgers and the first
higher--order equation of the Burgers hierarchy, which  can be
verified to be:
\begin{equation}
w = -k\left\{ 1 - \tanh \left[ kx - \left( 2k^2 - 4\epsilon
\alpha_4 k^3 \right) t \right] \right\} \quad .
\label{kkink}
\end{equation}

The second remark is about the more general transformation
alluded to above. Following ref.~\cite{fokas}, instead of
(\ref{phi}), we may alternatively introduce 
\begin{equation}
\phi \left( w \right) = \alpha w_x + \beta w^2 + \gamma w_x
\partial ^{-1}w + \nu x \left( w_{xx} + 2 ww_x \right) \quad ,
\label{newphi}
\end{equation}
with $\nu$ a constant.
This leads to the following normal form:
\begin{equation}
w_t = B\left( w \right) + \epsilon \left( \alpha_4 + 2\nu \right)
F_3\left( w \right)  +\epsilon \mu \left( \nu_1 w_x^2 + \nu_2
ww_{xx} + \nu_3 w^2 w_x \right)  \quad ,
\label{nformanormal}
\end{equation}
where $\mu $ is not modified, relation (\ref{eqnu}) is still
valid, but the  coefficients $\nu_1$ and $\nu_3$ have new
expressions in terms of the parameters  defining the
transformations. Explicitly, we have: 
\begin{equation}
\nu_1 = \mu^{-1} \left( 2\beta + 3\alpha_3 - 3 \alpha_4 -2\nu
\right) \quad ,
\end{equation}
\begin{equation}
\nu_3 = \mu^{-1} \left( 2\beta +\gamma + 3\alpha_1 - 3\alpha_4
-2\nu \right) \quad .
\end{equation}
The meaning of this result is the following: the new
transformation (\ref{newphi}) does  not have any influence on the
linearizability up to $\cal{O} (\epsilon)$ of 
Eq.(\ref{principal}), that is, it does not alter condition
(\ref{sol}). But, it makes it  possible to further simplify the
normal form by taking $\nu ={\alpha_4 }/{2}$. This completely
eliminates the $ F_3 \left( w \right) $ term from
Eq.(\ref{nformanormal}). 

\section*{Acknowledgement}

The authors would like to thank CNPq - Brazil, and FAPESP--Brazil
for  financial support.


\begin{thebibliography}{99}
\bibitem{dodd} R. K. Dodd, J. C. Eilbeck, J. D. Gibbon and H. C.
Morris, {\it Solitons and Nonlinear Wave Equations} (Academic,
London, 1982).
\bibitem{kpm1} R. A. Kraenkel, J. G. Pereira and M. A. Manna,
Phys. Rev. A 45 (1992) 45.
\bibitem{calo} F. Calogero, {\it Why Are Certain Nonlinear PDEs
Both Widely Applicable  and Integrable}, in  {\it What is
Integrability}, ed. by V. E. Zakharov (Springer, Berlin,  1991).
\bibitem{whit} G. B. Whitham, {\it Linear and Nonlinear Waves}
(Wiley, NY, 1973).
\bibitem{hier} D. V. Choodnovsky and G. V. Choodnovsky,  Nuov.
Cim. 40B (1977) 339.
\bibitem{arno} V. I. Arnold, {\it Geometrical Methods in the
Theory of Ordinary  Differential Equations} (Springer, Berlin,
1983). 
\bibitem{kodama1} Y. Kodama, Phys. Lett. 112A (1985) 193.
\bibitem{kodama2} Y. Kodama and A. V. Mikhailov, {\it Obstacles
to Asymptotic  Integrability}, in  {\it Algebraic Aspects of
Integrable Systems}, ed. by A. S.  Fokas and I. M. Gelfand
(Birkh\"auser, Basel, 1996).
\bibitem{bbm} R. A. Kraenkel, M. A. Manna, J. C. Montero and J. G.
Pereira, Phys. Rev. E (1996) 2976.  
\bibitem{fokas} A. S. Fokas and Q. M. Liu, Phys. Rev. Lett. 77
(1996) 2347.
\end{thebibliography}
\end{document}